\documentclass[twocolumn,showpacs,preprintnumbers,amsmath,amssymb]{revtex4}
\usepackage{amssymb}
\usepackage{graphicx}
\usepackage{dcolumn}
\usepackage{bm}
\begin{document}

\title{Unified parametrization of quark and lepton mixing matrices in tri-bimaximal pattern}

\author{Shi-Wen Li}
\affiliation{School of Physics, Peking University, Beijing 100871,
China}
\author{Bo-Qiang Ma\footnote{Corresponding author. Email address: \texttt{mabq@phy.pku.edu.cn}}}
\affiliation{School of Physics and State Key Laboratory of Nuclear
Physics and Technology, Peking University, Beijing 100871, China}

\begin{abstract}
Parametrization of the quark and lepton mixing matrices is the first
attempt to understand the mixing of fermions. In this work, we
parameterize the quark and lepton matrices with the help of
quark-lepton complementarity (QLC) in a tri-bimaximal pattern of
lepton mixing matrix. In this way, we combine the parametrization of
the two matrices with each other. We apply this new parametrization
to several physical quantities, and show its simplicity in the
expression of, e.g., the Jarlskog parameter of CP violation.
\end{abstract}


\maketitle

\section{Introduction}

Mixing of different generations of fermions is one of the most
important unsolved problems in particle physics, and has already
troubled physicists for half a century. Quark mixing of two
generations was first introduced by Cabibbo~\cite{cabibbo} decades
ago as a unified explanation to the difference of the coupling
constants in $\beta$ decay and $\Lambda$ decay in weak interaction.
This idea was extended to the case of three generations by Kobayashi
and Maskawa~\cite{km}. Parallelly, neutrino oscillation as the
lepton mixing was originally motivated by
Pontecorvo~\cite{pontecorvo} fifty years ago to understand the solar
neutrinoe deficit, and further generalized by Maki, Nakagawa and
Sakata~\cite{mns}. Visually, these mixings could be regarded as a
rotation from fermion mass eigenstates to flavor eigenstates.
Therefore, this rotation could be expressed as a mixing matrix,
namely the Cabibbo-Kobayashi-Maskawa (CKM) matrix $V$ for quark
sector and the Pontecorvo-Maki-Nakawaga-Sakata (PMNS) matrix $U$ for
lepton sector,
\begin{eqnarray}
V=\left(
  \begin{array}{ccc}
    V_{ud} & V_{us} & V_{ub} \\
    V_{cd} & V_{cs} & V_{cb} \\
    V_{td} & V_{ts} & V_{tb} \\
  \end{array}
\right), \quad
U=\left(
  \begin{array}{ccc}
    U_{e1}    & U_{e2}    & U_{e3}    \\
    U_{\mu1}  & U_{\mu2}  & U_{\mu3}  \\
    U_{\tau1} & U_{\tau2} & U_{\tau3} \\
  \end{array}
\right).\nonumber
\end{eqnarray}
Both the CKM and PMNS matrices are unitary matrices, comprising
three mixing angles $\theta_{ij}$ ($i,j=1,2,3$) and one Dirac
CP-violating phase $\delta$, and could thus be rewritten as
\begin{eqnarray}
\left(
\begin{array}{ccc}
c_{12}c_{31} & s_{12}c_{31} & s_{31}e^{-i\delta}           \\
-s_{12}c_{23}-c_{12}s_{23}s_{31}e^{i\delta} &
c_{12}c_{23}-s_{12}s_{23}s_{31}e^{i\delta}  & s_{23}c_{31} \\
s_{12}s_{23}-c_{12}c_{23}s_{31}e^{i\delta}  &
-c_{12}s_{23}-s_{12}c_{23}s_{31}e^{i\delta} & c_{23}c_{31}
\end{array}
\right),\nonumber
\end{eqnarray}
where $s_{ij}=\sin\theta_{ij}$ and $c_{ij}=\cos\theta_{ij}$. This
expression was first introduced by Chau and Keung~\cite{ck} in 1984,
and was advocated by the Particle Data Group in 1996. If neutrinos
are of Majorana type, there should be one additional diagonal matrix
with two Majorana phases ${\rm
diag}(e^{i\alpha_1/2},e^{i\alpha_2/2},1)$ multiplied to the matrix
above. However, the two Majorana phases do not affect neutrino
oscillations, and we ignore the diagonal matrix here.

The similarity of the CKM and PMNS matrices arouses a natural
question: what are their relations, e.g., the relations between
their elements, or equivalently, mixing angles? Is there a unified
comprehension of the two mixing matrices, just like the Cabibbo's
way out to interpret different decays with the same coupling
constant? Before the underlying grand unified theory of particle
physics is discovered, phenomenological analyses play the main role
in the correct direction. Amongst phenomenological analyses, to
parameterize the CKM and PMNS matrices may be the first attempt.
Parametrization provides us a numerical understanding of the
hierarchy of different elements in the mixing matrix, and this is
the purpose of our work.

For the CKM matrix, various particle physics experiments have
precisely fixed its value, with the allowed ranges of the magnitudes
of its elements reading~\cite{yao}
\begin{eqnarray}
\left(
  \begin{array}{ccc}
    0.97383^{+0.00024}_{-0.00023} & 0.2272^{+0.0010}_{-0.0010}    & 0.00396^{+0.00009}_{-0.00009} \\
    0.2271^{+0.0010}_{-0.0010}    & 0.97296^{+0.00024}_{-0.00024} & 0.04221^{+0.00010}_{-0.00080} \\
    0.00814^{+0.00032}_{-0.00064} & 0.04161^{+0.00012}_{-0.00078} & 0.999100^{+0.000034}_{-0.000004}
  \end{array} \right).\label{vn}
\end{eqnarray}
We find that the CKM matrix is approximately the unit matrix with
small deviations in the non-diagonal elements, and it could be
parameterized by the standard Wolfenstein
parametrization~\cite{wolfensteinpara},
\begin{eqnarray}
V=\left(
  \begin{array}{ccc}
    1-\frac{1}{2}\lambda^2   & \lambda                & A\lambda^3(\rho-i\eta) \\
    -\lambda                 & 1-\frac{1}{2}\lambda^2 & A\lambda^2             \\
    A\lambda^3(1-\rho-i\eta) & -A\lambda^2            & 1                      \\
  \end{array}
\right)+\mathcal{O}(\lambda^4),\label{wolfenstein}
\end{eqnarray}
with $\lambda=0.2272^{+0.0010}_{-0.0010}$ and
$A=0.818^{+0.007}_{-0.017}$~\cite{yao}. Nevertheless, the situation
of the PMNS matrix is exactly opposite. This is because (1) the
long-existed inaccuracy of neutrino experiments caused a delay in a
practical parametrization of the PMNS matrix as clear as the CKM
matrix; (2) almost all the elements in the PMNS matrix are of order
$\mathcal{O}(1)$ except $U_{e3}$, which makes the direct application
of the Wolfenstein-like perturbative expansion infeasible.

Whereas, circumstances have changed dramatically in recent years.
Thanks to more and more precise measurements of neutrino
oscillations, we have now a solid knowledge of the properties of
neutrinos, aspects of their mass squared differences and mixing
angles (although not as accurate as those of quarks). This helps us
to remove the first obstacle mentioned above, and in this paper, we
focus on the second one, i.e., how to parameterize the PMNS matrix
with large elements? Moreover, we try to find its relation with the
CKM matrix with the help of the so-called ``quark-lepton
complementarity'' (i.e., the sum of corresponding mixing angles in
quark and lepton sectors is about $45^{\circ}$, always referred as
QLC in literature), and parameterize both of them.

The organization of this paper is: in Sec. II, we take the
tri-bimaximal pattern as the basis of the PMNS matrix, but do not
parameterize the PMNS matrix around it immediately. However, we use
the QLC to obtain a new basis for the CKM matrix, and discuss its
parametrization. In Sec. III, with the new parameterized CKM matrix,
we utilize the QLC again, and the parametrization of the PMNS matrix
around the tri-bimaximal pattern is consequently derived, with the
same set of parameters in quark sector. Thus, we attain a unified
parametrization of both the CKM and PMNS matrices. Several physical
applications of the parametrization of the PMNS matrix are also
studied in detail. After that, we talk about the relations between
the two matrices and confirm the usefulness of the QLC when we
consider the parametrization of the PMNS matrix as an independent
one. Conclusions and discussions are summarized in Sec. IV.

\section{Parametrization of the CKM matrix}

Parametrization of the CKM matrix has already been well established,
i.e., the Wolfenstein parametrization in Eq.~(\ref{wolfenstein}),
which explicitly show the deviations of the non-diagonal elements
from the unit matrix at different powers of the parameter $\lambda$.
While, this method was presented long before the release of enormous
experimental data about neutrinos, and its parameters were
introduced, independently. Maybe a method of parametrization
connecting the quark sector to the lepton sector is more effective
at the epoch when fruitful neutrino experimental data are currently
available. Therefore, in this section, we contemplate a new method
for the parametrization of the CKM matrix, i.e., we parameterize it
around a new basis, which is obtained from the tri-bimaximal pattern
of the PMNS matrix by means of the QLC. Before continuing this
procedure, let us outline the main results of lepton mixing (i.e.,
neutrino oscillations).

Nowadays, vast experimental data have confirmed the oscillations
between different generations of neutrinos. For solar neutrino
oscillation, a global analysis of the KamLAND experiments yielded
$\Delta m^2_{21} = 7.9^{+0.6}_{-0.5} \times 10^{-5}~{\rm eV^2}$ and
$\tan^2\theta_{12} = 0.40^{+0.10}_{-0.07}$~\cite{kamland}; as for
SNO experiments, $\Delta m^2_{21} = 8.0^{+0.6}_{-0.4} \times
10^{-5}~{\rm eV^2}$ and $\theta_{12} =
(33.9^{+2.4}_{-2.2})^\circ$~\cite{sno}. Atmospheric neutrino
oscillation experimental data by different groups are: $1.5 \times
10^{-3}~{\rm eV^2} < |\Delta m^2_{23}| < 3.4 \times 10^{-3}~{\rm
eV^2}$ and $\sin^22\theta_{23} > 0.92$ at $90\%$ C.L.
(Super-Kamiokande Collaboration~\cite{kamiokande}); $|\Delta
m^2_{23}| = 2.74^{+0.44}_{-0.26} \times 10^{-3}~{\rm eV^2}$ and
$\sin^22\theta_{23} > 0.87$ at $68\%$ C.L. (MINOS
Collaboration~\cite{minos}); $1.9 \times 10^{-3}~{\rm eV^2} <
|\Delta m^2_{23}| < 3.5 \times 10^{-3}~{\rm eV^2}$ at $90\%$ C.L.
when $\sin^22\theta_{23} = 1$, with the best fit value of $2.8
\times 10^{-3}~{\rm eV^2}$ (K2K Collaboration~\cite{k2k}). On the
other hand, the CHOOZ experiment~\cite{chooz} indicated that the
upper bound of the mixing angle $\theta_{31}$ is pretty small. (See
Table \ref{tab}~\cite{maltoni} for a summary of the updated neutrino
oscillation parameters and also Ref.~\cite{murayama} for an
excellent illustration of the oscillation parameters.) According to
the results of global analysis, the modulus of the PMNS matrix is
summarized as (at $3\sigma$ C.L.)~\cite{gonzalez}
\begin{eqnarray}
|U|=\left(
  \begin{array}{ccc}
    0.77 - 0.86 & 0.50 - 0.63 & <0.22       \\
    0.22 - 0.56 & 0.44 - 0.73 & 0.57 - 0.80 \\
    0.21 - 0.55 & 0.40 - 0.71 & 0.59 - 0.82
  \end{array} \right),\label{u}
\end{eqnarray}

\begin{table} \centering
    \catcode`?=\active \def?{\hphantom{0}}
  \begin{tabular}
    {|@{\quad}>{\rule[-2mm]{0pt}{6mm}}l@{\quad}|@{\quad}c@{\quad}|@{\quad}c@{\quad}|@{\quad}c@{\quad}|}
    \hline
    parameters          & best fit & 2$\sigma$    & 3$\sigma$    \\
    \hline
    $\Delta m^2_{21}$   & 7.6??    & 7.3 - 8.1    & 7.1 - 8.3    \\
    $|\Delta m^2_{31}|$ & 2.4??    & 2.1 - 2.7    & 2.0 - 2.8    \\
    $\sin^2\theta_{12}$ & 0.32?    & 0.28 - 0.37  & 0.26 - 0.40  \\
    $\sin^2\theta_{23}$ & 0.50?    & 0.38 - 0.63  & 0.34 - 0.67  \\
    $\sin^2\theta_{31}$ & 0.007    & $\leq$ 0.033 & $\leq$ 0.050 \\
    \hline
  \end{tabular}
\caption{The three-flavor neutrino oscillation parameters from
global data including solar, atmospheric, reactor (KamLAND and
CHOOZ) and accelerator (K2K and MINOS) experiments. $\Delta
m^2_{21}$ is in units of $10^{-5}~{\rm eV^2}$ and $|\Delta
m^2_{31}|$ of $10^{-3}~{\rm eV^2}$.}\label{tab}
\end{table}

Eq.~(\ref{u}) manifests the nontriviality of the non-diagonal
elements in the PMNS matrix, and illuminates a distinct scheme for
the parametrization of the PMNS matrix. We should not apply the
method of the Wolfenstein parametrization of the CKM matrix directly
to the PMNS matrix, so firstly we need to find another basis for the
PMNS matrix. Otherwise, the parameter used for the expansion would
be of order $\mathcal{O}(1)$, the expansion of the PMNS matrix then
converges very slowly, and the hierarchy of different elements is
also rather vague.

The matrix chosen as the basis for our intention must be close to
the experimental constraints in Eq.~(\ref{u}). The most popular
basis matrices are the ``bimaximal" pattern (with
$\sin\theta_{12}=\sin\theta_{23}=\sqrt{2}/2$) and the
``tri-bimaximal" pattern (with $\sin\theta_{12}=\sqrt{3}/3$ and
$\sin\theta_{23}=\sqrt{2}/2$). The parametrization of the PMNS
matrix based on the bimaximal pattern was discussed
in~\cite{rodejohann}. In this paper, we concentrate on the
parametrization of the PMNS matrix around the tri-bimaximal pattern.

Interestingly, the tri-bimaximal pattern was also introduced by
Wolfenstein~\cite{wolfenstein2} thirty years ago with the form
\begin{eqnarray}
    U=\left(
        \begin{array}{ccc}
            \sqrt{6}/3  & \sqrt{3}/3  & 0          \\
           -\sqrt{6}/6  & \sqrt{3}/3  & \sqrt{2}/2 \\
            \sqrt{6}/6  & -\sqrt{3}/3 & \sqrt{2}/2
        \end{array}
        \right).\label{utribi}
\end{eqnarray}
Recently, this special matrix has received dense discussions in
literature~\cite{harrison}. Comparing Eq.~(\ref{u}) with
Eq.~(\ref{utribi}), we find that the tri-bimaximal pattern agrees
with the current experimental data fairly well, and the
parametrization around it must be reasonable.

Parametrization of the PMNS matrix around the tri-bimaximal pattern
was treated in several approaches~\cite{li}. Here, we do not repeat
these approaches, but postpone this parametrization in the next
section. Now, we want to see what happens to the CKM matrix if we
connect the tri-bimaximal pattern with the QLC.

QLC is an appealing relation between the quark and lepton mixing
angles,
\begin{eqnarray}
&&\theta_{12}^{\rm CKM}+\theta_{12}^{\rm PMNS}=45^{\circ},
\quad\theta_{23}^{\rm CKM}+\theta_{23}^{\rm PMNS}=45^{\circ},\nonumber\\
&&\theta_{31}^{\rm CKM}\approx\theta_{31}^{\rm PMNS}\approx
0^{\circ}.\label{qlc}
\end{eqnarray}
Below we use the superscripts $^{\rm CKM}$ and $^{\rm PMNS}$ to
denote the parameters of the CKM and PMNS matrices, respectively.
This relation was first introduced by Smirnov~\cite{smirnov}, and
has been heavily studied in different sorts of models~\cite{qlc}.
Eq.~(\ref{qlc}) is an extension for it.

Now that we have these empirical equations, the CKM and PMNS
matrices are not independent of each other, and therefore we are
enlightened to parameterize them in the same framework. With these
numerical equations and the tri-bimaximal pattern of the PMNS
matrix, we easily get the trigonometric functions of the mixing
angles in the CKM matrix,
\begin{eqnarray} &&\sin\theta_{12}^{\rm
CKM}=\frac{\sqrt{2}-1}{\sqrt{6}},
\quad \cos\theta_{12}^{\rm CKM}=\frac{\sqrt{2}+1}{\sqrt{6}},\nonumber\\
&&\sin\theta_{23}^{\rm CKM}=0, \quad \cos\theta_{23}^{\rm CKM}=1,\nonumber\\
&&\sin\theta_{31}^{\rm CKM}=0, \quad \cos\theta_{31}^{\rm
CKM}=1.\nonumber
\end{eqnarray}
The corresponding CKM matrix is thus fixed straightforwardly,
\begin{eqnarray}
    V=\left(
        \begin{array}{ccc}
            \frac{\sqrt{2}+1}{\sqrt{6}} & \frac{\sqrt{2}-1}{\sqrt{6}} & 0 \\
           -\frac{\sqrt{2}-1}{\sqrt{6}} & \frac{\sqrt{2}+1}{\sqrt{6}} & 0 \\
            0                           & 0                           & 1
        \end{array}
        \right).\label{v}
\end{eqnarray}

Eq.~(\ref{v}), being the lowest order approximation of the CKM
matrix, is not symmetric with $\theta_{12}^{\rm CKM}=9.7^\circ$,
$\theta_{23}^{\rm CKM}=0^\circ$ and $\theta_{31}^{\rm CKM}=0^\circ$.
Although this matrix is a little more complicated than the unit
matrix, it is closer to reality. If we take the bimaximal pattern of
the PMNS matrix with the QLC, we find that we get the unit matrix as
the lowest order approximation of the CKM matrix. In the following,
we just take Eq.~(\ref{v}) as the new basis for the CKM matrix, and
parameterize the CKM matrix around it.

Comparing Eq.~(\ref{v}) with Eq.~(\ref{vn}), we make an expansion of
$V$ in powers of $\lambda$ (Attention, $\lambda$ here is not the
Wolfenstein one, i.e., $V_{us}=\lambda$),
\begin{eqnarray}
V_{us}=\frac{\sqrt{2}-1}{\sqrt{6}}+\lambda,\nonumber
\end{eqnarray}
where $\lambda=0.0581^{+0.0010}_{-0.0010}$ measures the strength of
deviation of $V_{us}$ from the corresponding element in the matrix
we just obtained in Eq.~(\ref{v}). Since $0<\lambda<0.1$, this
expansion converges quickly. Similarly, we denote
\begin{eqnarray}
V_{cb}=A\lambda,\nonumber
\end{eqnarray}
and $A = 0.726^{+0.013}_{-0.018}$. As the CP violating phase is in
$V_{ub}$, and $V_{cb} \sim 10|V_{ub}|$, we may set
\begin{eqnarray}
V_{ub}=A\lambda^2(\rho-i\eta).\nonumber
\end{eqnarray}
Thus, $\sqrt{\rho^2+\eta^2} = 1.61^{+0.05}_{-0.06}$, and the values
of $\rho$ and $\eta$ correspond to the Dirac CP-violating phase
$\delta^{\rm CKM}$.

In summary, four parameters: $\lambda$, $A$, $\rho$ and $\eta$ are
introduced here,
\begin{eqnarray}
&&\lambda=0.0581^{+0.0010}_{-0.0010}, \quad A=0.726^{+0.013}_{-0.018},\nonumber\\
&&\sqrt{\rho^2+\eta^2}=1.61^{+0.05}_{-0.06}, \quad
1.23\lesssim\eta/\rho\lesssim4.70.\nonumber
\end{eqnarray}
where the ratio $\eta/\rho$ is fixed by
$\eta/\rho\simeq\tan\gamma\equiv-\tan\arg[(V_{ud}V_{ub}^{\ast})/(V_{cd}V_{cb}^{\ast})]$
($\gamma=(63^{+15}_{-12})^\circ$)~\cite{yao}, and these four
parameters characterize the CKM matrix completely.

Unitarity then determines the CKM matrix
\begin{widetext}
\begin{eqnarray}
    V&=&\left(
        \begin{array}{ccc}
            \frac{\sqrt{2}+1}{\sqrt{6}} & \frac{\sqrt{2}-1}{\sqrt{6}} & 0 \\
           -\frac{\sqrt{2}-1}{\sqrt{6}} & \frac{\sqrt{2}+1}{\sqrt{6}} & 0 \\
            0                           & 0                           & 1
        \end{array}
        \right)+\lambda\left(
        \begin{array}{ccc}
            -(3-2\sqrt{2})               & 1                             & 0 \\
            -1                           & -(3-2\sqrt{2})                & A \\
            \frac{\sqrt{2}-1}{\sqrt{6}}A & -\frac{\sqrt{2}+1}{\sqrt{6}}A & 0
        \end{array}\right)\nonumber\\
        &&+\lambda^2\left(
        \begin{array}{ccc}
            -(30\sqrt{3}-21\sqrt{6})        & 0                                 & (\rho-i\eta)A \\
            \frac{\sqrt{2}-1}{2\sqrt{6}}A^2 &-(30\sqrt{3}-21\sqrt{6})-\frac{\sqrt{2}+1}{2\sqrt{6}}A^2 & 0 \\
            \left(1-\frac{\sqrt{2}+1}{\sqrt{6}}(\rho+i\eta)\right)A
            & \left(3-2\sqrt{2}-\frac{\sqrt{2}-1}{\sqrt{6}}(\rho+i\eta)\right)A & -{1\over2}A^2
        \end{array}\right)+\mathcal{O}(\lambda^3).\label{ve}
\end{eqnarray}
\end{widetext}

We can extract the information of quark mixing from Eq.~(\ref{ve}):
\begin{enumerate}
\item
The expansion is reasonable in powers of $\lambda$ and converges
quickly.

\item
The term of $\lambda^0$ is the new basis we have introduced in
Eq.~(\ref{v}).

\item
The term of $\lambda^1$ clearly shows the deviation of the CKM
matrix from this new basis, and it contains only one parameter $A$
except $\lambda$.

\item
The term of $\lambda^2$ is the modification of higher order, and the
effect of CP violation exists in this order. Since CP violation is
stored in $V_{ub}$, the degree of CP violation in quark sector is of
order $\lambda^2$ in our parametrization.
\end{enumerate}

In this section, the CKM matrix is parameterized in a new way, i.e.,
around a new matrix, which is considered as the basis of the CKM
matrix. The crucial input is the QLC. Following this idea, we make
use of the QLC once more, and the parametrization of the PMNS matrix
is then obtained.

\section{Parametrization of the PMNS matrix}

Although many properties of neutrinos have been known from neutrino
oscillation experiments, information is not enough to ensure the
accurate ranges of the modulus of the elements in the PMNS matrix.
Known from Eq.~(\ref{u}), the parametrization of the PMNS matrix is
troublesome if we do not take a basis matrix into account. So we
need a method to deal with it and make the parametrization feasible.
Fortunately, the QLC connects the mixing angles of the two matrices
which indicates that we may get the PMNS matrix from the CKM matrix.
In this paper, we do not take the Wolfenstein parametrization of the
CKM matrix as the perturbative expansion which leads to another
parametrization of the PMNS matrix with the bimaximal pattern
(see~\cite{li2} for details), instead, we use Eq.~(\ref{ve}) we have
just obtained for the CKM matrix. With the QLC, we find the
trigonometric functions of the mixing angles in the PMNS matrix (to
order of $\lambda^2$),
\begin{eqnarray} &&\sin\theta_{12}^{\rm
PMNS}=\frac{\sqrt{3}}{3}-(2\sqrt{2}-2)\lambda-(15\sqrt{6}-21\sqrt{3})\lambda^2,\nonumber\\
&&\cos\theta_{12}^{\rm PMNS}=\frac{\sqrt{6}}{3}+(2-\sqrt{2})\lambda-(15\sqrt{6}-21\sqrt{3})\lambda^2,\nonumber\\
&&\sin\theta_{23}^{\rm PMNS}=\frac{\sqrt{2}}{2}\left(1-A\lambda-\frac{1}{2}A^2\lambda^2\right),\nonumber\\
&&\cos\theta_{23}^{\rm PMNS}=\frac{\sqrt{2}}{2}\left(1+A\lambda-\frac{1}{2}A^2\lambda^2\right),\nonumber\\
&&\sin\theta_{31}^{\rm PMNS}=A\lambda^2\sqrt{\zeta^2+\xi^2},\nonumber\\
&&\cos\theta_{31}^{\rm PMNS}=1,\label{tri}
\end{eqnarray}
where $A$ and $\lambda$ are the parameters we have defined for the
CKM matrix, i.e., we describe the CKM and PMNS matrices with the
same set of parameters. As we see in the parametrization of the CKM
matrix, there are totally four parameters, but from Eq.~(\ref{qlc})
we only know two precise numerical equations. It implies that we
have to introduce another two new parameters $\zeta$ and $\xi$ to
describe the PMNS matrix completely. From the fifth equation in
Eq.~(\ref{tri}), $\sin\theta_{31}^{\rm PMNS}e^{-i\delta^{\rm
PMNS}}=A\lambda^2(\zeta-i\xi)= A\lambda^2z^\ast$, where
$z\equiv\zeta+i\xi$. According to the experimental data of neutrino
oscillations, the value of $\sin\theta_{31}^{\rm PMNS}$ has not been
fixed yet, and what we have known is only its upper bound (see
Table~\ref{tab}). We find that $|U_{e3}|=\sin\theta_{31}^{\rm
PMNS}<0.22$ at $3\sigma$ C.L. from Eq.~(\ref{u}). If $0.22$ is taken
for the value of $|U_{e3}|$, we find that $\sqrt{\zeta^2+\xi^2} \sim
90$, which is unsuitable for the expansion of the PMNS matrix
because the hierarchy of the first three orders would be indefinite.
But if the value of $|U_{e3}|$ is smaller than $0.01$,
$\sqrt{\zeta^2+\xi^2}\lesssim4$ and $\sin\theta_{31}^{\rm
PMNS}e^{-i\delta^{\rm PMNS}}=A\lambda^2z^\ast$ is preferable. The
value of $|U_{e3}|$ depends on the experimental data, hence we may
also set $\sin\theta_{31}^{\rm PMNS}e^{-i\delta^{\rm
PMNS}}=A\lambda(\zeta'-i\xi')=A\lambda z'^\ast$
($\xi'/\zeta'=\xi/\zeta=\tan\delta^{\rm PMNS}$), where
$z'^\ast\equiv\zeta'+i\xi'$. $\sqrt{\zeta'^2+\xi'^2} \sim 5$ if we
take $0.22$ for the value of $|U_{e3}|$, which means that
$\sin\theta_{31}^{\rm PMNS}e^{-i\delta^{\rm PMNS}}=A\lambda z'^\ast$
would be better if the value of $|U_{e3}|$ is not small enough. Let
us discuss the difference between the two cases.

\begin{widetext}
{\it Case 1}: $\sin\theta_{31}^{\rm PMNS}e^{-i\delta^{\rm
PMNS}}=A\lambda^2z^\ast$.

With the trigonometric functions of the mixing angles in the PMNS
matrix we have the parametrization of the PMNS matrix
\begin{eqnarray}
U&=&\left(
  \begin{array}{ccc}
    \frac{\sqrt{6}}{3}  & \frac{\sqrt{3}}{3} & 0                  \\
    -\frac{\sqrt{6}}{6} & \frac{\sqrt{3}}{3} & \frac{\sqrt{2}}{2} \\
    \frac{\sqrt{6}}{6}  & -\frac{\sqrt{3}}{3}& \frac{\sqrt{2}}{2}
  \end{array}
\right)+\lambda
\left(
  \begin{array}{ccc}
    2-\sqrt{2}                        & -(2\sqrt{2}-2)                    & 0                    \\
    2-\sqrt{2}-\frac{1}{\sqrt{6}}A    & \sqrt{2}-1+\frac{1}{\sqrt{3}}A    & -\frac{\sqrt{2}}{2}A \\
    -(2-\sqrt{2})-\frac{1}{\sqrt{6}}A & -(\sqrt{2}-1)+\frac{1}{\sqrt{3}}A & \frac{\sqrt{2}}{2}A
  \end{array}
\right)\nonumber\\
&&+\lambda^2\left(
  \begin{array}{ccc}
    -(15\sqrt{6}-21\sqrt{3}) & -(15\sqrt{6}-21\sqrt{3}) & z^\ast A \\
    15\sqrt{3}-\frac{63}{\sqrt{6}}+\left(2-\sqrt{2}-\frac{z}{\sqrt{3}}\right)A+\frac{1}{2\sqrt{6}}A^2
    & -\left(15\sqrt{3}-\frac{63}{\sqrt{6}}\right)+\left(\sqrt{2}-1-\frac{z}{\sqrt{6}}\right)A
    -\frac{1}{2\sqrt{3}}A^2  & -\frac{\sqrt{2}}{4}A^2 \\
    -\left(15\sqrt{3}-\frac{63}{\sqrt{6}}\right)+\left(2-\sqrt{2}-\frac{z}{\sqrt{3}}\right)A-\frac{1}{2\sqrt{6}}A^2
    & 15\sqrt{3}-\frac{63}{\sqrt{6}}+\left(\sqrt{2}-1-\frac{z}{\sqrt{6}}\right)A
    +\frac{1}{2\sqrt{3}}A^2  & -\frac{\sqrt{2}}{4}A^2
  \end{array}
\right)\nonumber\\
&&+\mathcal{O}(\lambda^3).\label{ue2}
\end{eqnarray}

{\it Case 2}: $\sin\theta_{31}^{\rm PMNS}e^{-i\delta^{\rm
PMNS}}=A\lambda z'^\ast$.

In this case, the last two equations in Eq.~(\ref{tri}) are replaced
by (to order of $\lambda^2$)
\begin{eqnarray}
\sin\theta_{31}^{\rm PMNS}=A\lambda\sqrt{(\zeta'^2+\xi'^2)}, \quad
\cos\theta_{31}^{\rm
PMNS}=1-{1\over2}A^2\lambda^2(\zeta'^2+\xi'^2).\nonumber
\end{eqnarray}
Repeating the former process, we have
\begin{eqnarray}
U&=&\left(
  \begin{array}{ccc}
    \frac{\sqrt{6}}{3}  & \frac{\sqrt{3}}{3}  & 0                  \\
    -\frac{\sqrt{6}}{6} & \frac{\sqrt{3}}{3}  & \frac{\sqrt{2}}{2} \\
    \frac{\sqrt{6}}{6}  & -\frac{\sqrt{3}}{3} & \frac{\sqrt{2}}{2}
  \end{array}
\right)+\lambda
\left(
  \begin{array}{ccc}
    2-\sqrt{2} & -(2\sqrt{2}-2) & z'^\ast A \\
    2-\sqrt{2}-\left(\frac{1}{\sqrt{6}}+\frac{z'}{\sqrt{3}}\right)A
    & \sqrt{2}-1+\left(\frac{1}{\sqrt{3}}-\frac{z'}{\sqrt{6}}\right)A    & -\frac{\sqrt{2}}{2}A \\
    -(2-\sqrt{2})-\left(\frac{1}{\sqrt{6}}+\frac{z'}{\sqrt{3}}\right)A
    & -(\sqrt{2}-1)+\left(\frac{1}{\sqrt{3}}-\frac{z'}{\sqrt{6}}\right)A & \frac{\sqrt{2}}{2}A
  \end{array}
\right)\nonumber\\
&&+\lambda^2\left(
  \begin{array}{ccc}
    -(15\sqrt{6}-21\sqrt{3})-\frac{\sqrt{6}}{6}|z'|^2A^2 & -(15\sqrt{6}-21\sqrt{3})-\frac{\sqrt{3}}{6}|z'|^2A^2 & 0  \\
    15\sqrt{3}-\frac{63}{\sqrt{6}}+mA+pA^2               & -\left(15\sqrt{3}-\frac{63}{\sqrt{6}}\right)+nA-qA^2
    & -\frac{\sqrt{2}}{4}(1+|z'|^2)A^2 \\
    -\left(15\sqrt{3}-\frac{63}{\sqrt{6}}\right)+mA-pA^2 & 15\sqrt{3}-\frac{63}{\sqrt{6}}+nA+qA^2
    & -\frac{\sqrt{2}}{4}(1+|z'|^2)A^2
  \end{array}
\right)+\mathcal{O}(\lambda^3),\label{ue1}
\end{eqnarray}
where
\begin{eqnarray}
m=2-\sqrt{2}-\frac{z'}{\sqrt{2}+1},\quad
n=\sqrt{2}-1+\frac{2z'}{2+\sqrt{2}},\quad
p=\frac{1}{2\sqrt{6}}+\frac{z'}{\sqrt{3}},\quad
q=\frac{1}{2\sqrt{3}}-\frac{z'}{\sqrt{6}}.\nonumber
\end{eqnarray}
\end{widetext}

Now we see the meanings of the expansions of $U$ in Eq.~(\ref{ue2})
and Eq.~(\ref{ue1}):
\begin{enumerate}
\item
The PMNS matrix $U$ is expanded in powers of $\lambda$ which makes
the expansion converge quickly, the same as the CKM matrix $V$. So
the parameter $\lambda$ we choose for the CKM matrix is quite well.

\item
The term of $\lambda^0$ is the zeroth order approximation, we find
that it is just the tri-bimaximal pattern of the PMNS matrix, i.e.,
the matrix we want to take as the basis of the PMNS matrix. So the
method of getting the expansion of the PMNS matrix from the
expansion of the CKM matrix with the QLC is feasible.

\item
The term of $\lambda^1$ shows the deviation of the PMNS matrix from
the tri-bimaximal pattern. We find that the deviation is real and
just relates with $\lambda$ and $A$ in Eq. (\ref{ue2}). On the other
hand, it is complex and relates with all four parameters in
Eq.~(\ref{ue1}).

\item
The term of $\lambda^2$ is the modification of higher order, and it
is much more complicated in Eq.~(\ref{ue1}) than Eq.~(\ref{ue2}).

\item
CP violation is described by $U_{e3}$ in the PMNS matrix, and we
find that the degree of CP violation in lepton sector is of order
$\lambda^2$ in the parametrization of {\it case 1} but of order
$\lambda^1$ in that of {\it case 2}. Since the choice of {\it case
1} or {\it case 2} depends on the value of $|U_{e3}|$, it implies
that the size of CP violation is determined by the magnitude of
$\sin\theta_{31}^{\rm PMNS}$.
\end{enumerate}

Now let us apply the parametrization of the PMNS matrix to some
physical observables.

The amplitude for neutrinoless double beta decay ($0\nu\beta\beta$)
is proportional to the quantity~\cite{yao}
\begin{eqnarray}
|\langle m_{\beta\beta}\rangle|\equiv\left|\sum_i
U_{ei}^2m_i\right|.\nonumber
\end{eqnarray}
Since the two Majorana CP-violating phases are related here, we must
take them into account.

In {\it case 1} (to order of $\lambda^2$),
\begin{eqnarray}
|\langle
m_{\beta\beta}\rangle_1|&=&\bigg|\frac{1}{3}(2m_1e^{i\alpha_1}+m_2e^{i\alpha_2})\nonumber\\
&&+\frac{4}{3}(\sqrt{6}-\sqrt{3})\lambda(m_1e^{i\alpha_1}-m_2e^{i\alpha_2})\nonumber\\
&&-(54-38\sqrt{2})\lambda^2(m_1e^{i\alpha_1}-m_2e^{i\alpha_2})\bigg|.\nonumber
\end{eqnarray}
We find that only $m_1$ and $m_2$ affect the value of $|\langle
m_{\beta\beta}\rangle_1|$, $m_3$ vanishes here to order of
$\lambda^2$. Furthermore, there is only one parameter $\lambda$
which in our parametrization shows up in $|\langle
m_{\beta\beta}\rangle_1|$.

In {\it case 2} (to order of $\lambda^2$),
\begin{eqnarray}
|\langle m_{\beta\beta}\rangle_2|
&=&\bigg|\langle m_{\beta\beta}\rangle_1+A^2\lambda^2\bigg((\zeta'-i\xi')^2m_3\nonumber\\
&&-\frac{1}{3}(\zeta'^2+\xi'^2)(2m_1e^{i\alpha_1}+m_2e^{i\alpha_2})\bigg)\bigg|.\nonumber
\end{eqnarray}
Apparently, $|\langle m_{\beta\beta}\rangle_2|$ is more complicated
than $|\langle m_{\beta\beta}\rangle_1|$. Though the four parameters
in our parametrization all appear in $|\langle
m_{\beta\beta}\rangle_2|$, the effect of $m_3$ is still small,
suppressed by $\lambda^2$.

The rephasing-invariant Jarlskog parameter
$\mathcal{J}$~\cite{jarlskog} of CP violation is given by
\begin{eqnarray}
\mathcal{J}\equiv{\rm Im}(U_{e2}U_{\mu3}U_{e3}^\ast U_{\mu2}^\ast)
=s_{12}s_{23}s_{31}c_{12}c_{23}c_{31}^2\sin\delta.\nonumber
\end{eqnarray}
An essential feature of $\mathcal{J}$ is that it is phase-convention
independent which makes it very important when discussing CP
violation. The parameter is expressed in a simple form in our
parametrization (to order of $\lambda^2$). In {\it case 1}, we have
\begin{eqnarray}
\mathcal{J}=\frac{\sqrt{2}}{6}A\lambda^2\xi,\nonumber
\end{eqnarray}
and in {\it case 2}, we have
\begin{eqnarray}
\mathcal{J}=\left(\frac{\sqrt{2}}{6}-\frac{\sqrt{2}-1}{\sqrt{6}}\lambda\right)A\lambda\xi'.\nonumber
\end{eqnarray}
If we rescale $\zeta'\rightarrow\lambda\zeta$ and
$\xi'\rightarrow\lambda\xi$, we will find that $\mathcal{J}$ in {\it
case 2} becomes equivalent to $\mathcal{J}$ in {\it case 1} (to
order of $\lambda^2$). If we take $\sin^2\theta_{31}^{\rm
PMNS}=0.007$ as a preliminary estimate,
$\sqrt{\zeta'^2+\xi'^2}\sim2$, then $\zeta'$ and $\xi'$ are of order
$\mathcal {O}(1)$, {\it case 2} is preferable. However, if
$\sin^2\theta_{31}^{\rm PMNS}=0.0001$ or less, $\zeta$ and $\xi$ are
of order $\mathcal {O}(1)$, then {\it case 1} is preferred.

Leptonic unitarity triangles~\cite{farzan} may be constructed with
the orthogonality of different pairs of rows or columns of the PMNS
matrix. The matrix elements $U_{e2}$, $U_{e3}$ and $U_{\mu3}$ play
important roles in the experiments and they have relatively simple
forms in our parametrization. There are two unitarity triangles
which contain the three matrix elements, the ``$\nu_2,\nu_3$"
triangle and the ``$\nu_e,\nu_\mu$" triangle. Here we take the
``$\nu_2,\nu_3$" unitarity triangle as an example (for previous
discussions see Bjorken and King in~\cite{li}).

The unitarity relation of the ``$\nu_2,\nu_3$" triangle is
\begin{eqnarray}
U_{e2}U_{e3}^\ast+U_{\mu2}U_{\mu3}^\ast+U_{\tau2}U_{\tau3}^\ast=0.\label{ur}
\end{eqnarray}
\begin{widetext}
In {\it case 1} (to order of $\lambda^2$), we have
\begin{eqnarray}
&&U_{e2}U_{e3}^\ast=\frac{z}{\sqrt{3}}A\lambda^2,\nonumber\\
&&U_{\mu2}U_{\mu3}^\ast=\frac{1}{\sqrt{6}}+\left(1-\frac{1}{\sqrt{2}}\right)\lambda
-\left(\frac{45}{\sqrt{6}}-\frac{63}{\sqrt{12}}+\frac{z}{2\sqrt{3}}A+\frac{\sqrt{6}}{3}A^2\right)\lambda^2,\nonumber\\
&&U_{\tau2}U_{\tau3}^\ast=-\frac{1}{\sqrt{6}}-\left(1-\frac{1}{\sqrt{2}}\right)\lambda
+\left(\frac{45}{\sqrt{6}}-\frac{63}{\sqrt{12}}-\frac{z}{2\sqrt{3}}A+\frac{\sqrt{6}}{3}A^2\right)\lambda^2,\nonumber
\end{eqnarray}
and in {\it case 2} (to order of $\lambda^2$),
\begin{eqnarray}
&&U_{e2}U_{e3}^\ast=\frac{z'}{\sqrt{3}}A\lambda
-\frac{2z'}{\sqrt{2}+1}A\lambda^2,\nonumber\\
&&U_{\mu2}U_{\mu3}^\ast=\frac{1}{\sqrt{6}}+\left(1-\frac{1}{\sqrt{2}}-\frac{z'}{2\sqrt{3}}A\right)\lambda
-\Bigg(\frac{45}{\sqrt{6}}-\frac{63}{\sqrt{12}}-\frac{z'}{\sqrt{2}+1}A
+\left(\frac{\sqrt{6}}{3}-\frac{z'}{\sqrt{3}}+\frac{|z'|^2}{2\sqrt{6}}\right)A^2\Bigg)\lambda^2,\nonumber\\
&&U_{\tau2}U_{\tau3}^\ast=-\frac{1}{\sqrt{6}}-\left(1-\frac{1}{\sqrt{2}}+\frac{z'}{2\sqrt{3}}A\right)\lambda
+\Bigg(\frac{45}{\sqrt{6}}-\frac{63}{\sqrt{12}}+\frac{z'}{\sqrt{2}+1}A
+\left(\frac{\sqrt{6}}{3}-\frac{z'}{\sqrt{3}}+\frac{|z'|^2}{2\sqrt{6}}\right)A^2\Bigg)\lambda^2.\nonumber
\end{eqnarray}

We find that the $U_{e2}U_{e3}^\ast$ side of the ``$\nu_2,\nu_3$"
unitarity triangle is very small, suppressed by $\lambda^2$ in {\it
case 1} and $\lambda$ in {\it case 2}. This is because $|U_{e2}|$ is
of $\mathcal{O}(1)$, and $|U_{e3}|$ is of $\mathcal{O}(\lambda^2)$
in {\it case 1} but $\mathcal{O}(\lambda)$ in {\it case 2}.
Obviously, Eq.~(\ref{ur}) has been satisfied in both cases (to order
of $\lambda^2$). If neutrinos are of Majorana type, there should be
one additional factor $e^{i\alpha_2/2}$ multiplied to every
expression, but Eq.~(\ref{ur}) still holds.

At last, we turn to neutrino oscillations. The probability of the
oscillation from flavor $\nu_{\alpha}$ to flavor $\nu_{\beta}$,
$P(\nu_\alpha\rightarrow\nu_\beta)$ is given by
\begin{eqnarray}
P(\nu_\alpha\rightarrow\nu_\beta)&=&\left|\sum_{i=1}^3U_{\alpha
i}^\ast e^{-im_i^2\frac{L}{2E}}U_{\beta i}\right|^2\nonumber\\
&&=\delta_{\alpha\beta}-4\sum_{i>j}{\rm Re}(U_{\alpha i}^\ast
U_{\beta i}U_{\alpha j}U_{\beta
j}^\ast)\sin^2\Delta_{ij}+2\sum_{i>j}{\rm Im}(U_{\alpha i}^\ast
U_{\beta i}U_{\alpha j}U_{\beta j}^\ast)\sin2\Delta_{ij},\nonumber
\end{eqnarray}
where $\alpha$, $\beta=e$, $\mu$, $\tau$,
$\Delta_{ij}\equiv(m_i^2-m_j^2)L/4E$, $L$ is the oscillation length
and $E$ is the beam energy of neutrinos.

Let us take $P(\nu_e\rightarrow\nu_e)$ for example (for other
oscillation probabilities see the Appendix). In {\it case 1} (to
order of $\lambda^2$), we obtain
\begin{eqnarray}
P(\nu_e\rightarrow\nu_e)&=&1-\left(\frac{8}{9}-\frac{16}{9}(\sqrt{6}-\sqrt{3})\lambda-8(\sqrt{2}-1)\lambda^2\right)
\sin^2\Delta_{21},\nonumber
\end{eqnarray}
and in {\it case 2} (to order of $\lambda^2$), we have
\begin{eqnarray}
P(\nu_e\rightarrow\nu_e)&=&1-\left(\frac{8}{9}-\frac{16}{9}(\sqrt{6}-\sqrt{3})\lambda-
\left(8(\sqrt{2}-1)+\frac{16}{9}(\zeta'^2+\xi'^2)A^2\right)\lambda^2\right)\sin^2\Delta_{21}\nonumber\\
&&-\frac{4}{3}(\zeta'^2+\xi'^2)A^2\lambda^2(2\sin^2\Delta_{31}+\sin^2\Delta_{32}).\nonumber
\end{eqnarray}
\end{widetext}
Certainly we find that $\sum_\beta
P(\nu_\alpha\rightarrow\nu_\beta)=1$ to order of $\lambda^2$, which
means that the sum of the probability that a neutrino changes its
flavor and the probability that it does not is unity.

Nevertheless, in despite of the CKM matrix and the QLC, we may
consider Eq.~(\ref{tri}) as an independent parametrization of the
PMNS matrix based on the tri-bimaximal pattern. Then Eq.~(\ref{ue2})
and Eq.~(\ref{ue1}) are still obtained. According to the data in
Table \ref{tab}, we find the ranges of the parameters in the PMNS
matrix (at $3\sigma$ C.L.),
\begin{eqnarray}
&&-0.069<\lambda^{\rm PMNS}<0.079,\nonumber\\
&&-0.17<A^{\rm PMNS}\lambda^{\rm PMNS}<0.16,\nonumber
\end{eqnarray}
and $\sqrt{\zeta^2+\xi^2}<16$ in {\it case 1},
$\sqrt{\zeta'^2+\xi'^2}<1.3$ in {\it case 2}.
\newpage
Without loss of
generality, we set $A^{\rm PMNS}>0$, and we only acquire the upper
bound of $A^{\rm PMNS}$,
\begin{eqnarray}
A^{\rm PMNS}<2.5.\nonumber
\end{eqnarray}
Clearly, the ranges of $\lambda^{\rm CKM}$ and $A^{\rm CKM}$ are
included in the ranges of $\lambda^{\rm PMNS}$ and $A^{\rm PMNS}$
since the experiments for the PMNS matrix are not as accurate as
those for the CKM matrix. It implies that the corresponding
parameters of the two matrices may be correlated with each other to
some extent when the parametrization of the PMNS matrix is
independent. Let us see if there are any relations between the
parameters of the CKM matrix and those of the PMNS matrix.
\begin{widetext}
\begin{eqnarray}
\sin(\theta^{\rm CKM}_{12}+\theta^{\rm PMNS}_{12})
&=&\frac{\sqrt{2}}{2}+(\sqrt{6}-\sqrt{3})(\lambda^{\rm
CKM}-\lambda^{\rm PMNS}) -(30-21\sqrt{2})(\lambda^{\rm CKM})^2
-(30\sqrt{2}-42)(\lambda^{\rm PMNS})^2\nonumber\\
&&+(9\sqrt{2}-12)\lambda^{\rm CKM}\lambda^{\rm PMNS},\nonumber\\
\sin(\theta^{\rm CKM}_{23}+\theta^{\rm PMNS}_{23})
&=&\frac{\sqrt{2}}{2}+\frac{\sqrt{2}}{2}(A^{\rm CKM}\lambda^{\rm
CKM} -A^{\rm PMNS}\lambda^{\rm PMNS}) -\frac{\sqrt{2}}{4}(A^{\rm
CKM}\lambda^{\rm CKM}-A^{\rm PMNS}\lambda^{\rm PMNS})^2.\nonumber
\end{eqnarray}
\end{widetext}
It is clear that the QLC is reobtained to zeroth order, and the two
equations above coincide with the first two equations in
Eq.~(\ref{qlc}) to order of $\lambda^2$ if $\lambda^{\rm
CKM}=\lambda^{\rm PMNS}$ and $A^{\rm CKM}=A^{\rm PMNS}$. If the
relation between $\theta_{31}^{\rm CKM}$ and $\theta_{31}^{\rm
PMNS}$ is explicit, there should be an equation which links
$\sqrt{\rho^2+\eta^2}$ to $\sqrt{\zeta^2+\xi^2}$, where the values
of $\rho$ and $\eta$ correspond to $\delta^{\rm CKM}$, and the
values of $\zeta$ and $\xi$ to $\delta^{\rm PMNS}$. Thus, we state
that the parameters in the PMNS matrix are not independent although
the parametrization of the PMNS matrix is independent. The relations
between the parameters of the CKM and PMNS matrices could be
realized by the QLC. On another point of view, we may consider the
QLC as the zeroth order approximation to the sum of the
corresponding mixing angles of the CKM and PMNS matrices.

The expansion of the PMNS matrix has been derived directly based on
the tri-bimaximal pattern of the PMNS matrix in another work (see
the first paper in~\cite{li2}). In that case, the parametrization is
independent, and the expansion is reasonable in powers of $\lambda$
($|\lambda|<0.1$ at $3\sigma$ C.L.). The degree of CP violation is
of order $\lambda^1$, the Jarlskog parameter is combined of two
terms, and the effect of $m_3$ is suppressed by $\lambda^2$ in the
amplitude of neutrinoless double beta decay. It means that the
expansion in that case is closer to {\it case 2} in this paper where
the parameters in the CKM and PMNS matrices are not independent no
matter the parametrization of the PMNS matrix is independent or not.
The QLC plays an important role in the combination of the two
matrices. Nevertheless, the magnitude of $\theta_{31}^{\rm PMNS}$ is
not explicit, hence the parametrization of the PMNS matrix is
arbitrary to some extent and the choice between Eq.~(\ref{ue2}) and
Eq.~(\ref{ue1}) depends on the experimental data of $|U_{e3}|$.

\section{Summary}

In this work, we introduce a new basis for the CKM matrix, which is
closer to reality than the unit matrix. The unit matrix is symmetric
with no mixing angles while the new basis matrix is not symmetric
with one non-vanishing mixing angle. Taking the tri-bimaximal
pattern as the zeroth approximation for the PMNS matrix, the new
basis of the CKM matrix is obtained via the QLC. We parameterize the
CKM matrix around the new basis instead of the unit matrix, and find
the expansion is suitable as a perturbative approximation.

The parametrization of the PMNS matrix is attained from the
expansion of the CKM matrix with the QLC, then the two matrices are
unified with the same set of parameters. Nevertheless, we get the
expansion of the PMNS matrix, even if we consider parametrization in
lepton sector as an independent one. In this case, however, the
parameters of the CKM matrix are still associated with those of the
PMNS matrix, since the QLC is reobtained when we combine the
parameters of the two matrices.

Anyway, the expansions of the CKM and PMNS matrices are reasonable
and converge quickly since the parameter $\lambda$ is very small.
The effect of CP violation is also seen from the parametrization of
the two matrices. Furthermore, some observable quantities are
expressed in relatively simple forms or only related with one or two
parameters in our parametrization. Finally, the two matrices are
correlated with each other and useful for analyzing experimental
data and phenomenology in particle physics.

\begin{acknowledgments}
We are grateful to Nan Li for valuable contributions and to Zhiqiang
Guo for discussions. This work is partially supported by National
Natural Science Foundation of China (Nos.~10721063, 10575003,
10528510), by the Key Grant Project of Chinese Ministry of Education
(No.~305001), and by the Research Fund for the Doctoral Program of
Higher Education (China).

\end{acknowledgments}

\begin{widetext}
\begin{center}
{\bf APPENDIX}
\end{center}

In the appendix, we list all other probabilities for neutrino
oscillations (to order of $\lambda^2$).

In {\it case 1},
\begin{eqnarray}
P(\nu_e\rightarrow\nu_\mu)&=&\left(\frac{4}{9}-\frac{8}{9}(\sqrt{6}-\sqrt{3}-A)\lambda-\left(4(\sqrt{2}-1)
+\frac{16}{9}\left(\sqrt{6}-\sqrt{3}-\frac{\sqrt{2}}{8}\zeta\right)A\right)\lambda^2\right)\sin^2\Delta_{21}\nonumber\\
&&+\frac{2\sqrt{2}}{3}\zeta
A\lambda^2(\sin^2\Delta_{31}-\sin^2\Delta_{32})
+\frac{\sqrt{2}}{3}\xi
A\lambda^2(\sin2\Delta_{21}-\sin2\Delta_{31}+\sin2\Delta_{32}),\nonumber\\
P(\nu_e\rightarrow\nu_\tau)&=&\left(\frac{4}{9}-\frac{8}{9}(\sqrt{6}-\sqrt{3}+A)\lambda-\left(4(\sqrt{2}-1)
-\frac{16}{9}\left(\sqrt{6}-\sqrt{3}-\frac{\sqrt{2}}{8}\zeta\right)A\right)\lambda^2\right)\sin^2\Delta_{21}\nonumber\\
&&-\frac{2\sqrt{2}}{3}\zeta
A\lambda^2(\sin^2\Delta_{31}-\sin^2\Delta_{32})
-\frac{\sqrt{2}}{3}\xi
A\lambda^2(\sin2\Delta_{21}-\sin2\Delta_{31}+\sin2\Delta_{32}),\nonumber\\
P(\nu_\mu\rightarrow\nu_e)&=&\left(\frac{4}{9}-\frac{8}{9}(\sqrt{6}-\sqrt{3}-A)\lambda-\left(4(\sqrt{2}-1)
+\frac{16}{9}\left(\sqrt{6}-\sqrt{3}-\frac{\sqrt{2}}{8}\zeta\right)A\right)\lambda^2\right)\sin^2\Delta_{21}\nonumber\\
&&+\frac{2\sqrt{2}}{3}\zeta
A\lambda^2(\sin^2\Delta_{31}-\sin^2\Delta_{32})
-\frac{\sqrt{2}}{3}\xi
A\lambda^2(\sin2\Delta_{21}-\sin2\Delta_{31}+\sin2\Delta_{32}),\nonumber\\
P(\nu_\mu\rightarrow\nu_\mu)&=&1-\left(\frac{2}{9}-\frac{4}{9}(\sqrt{6}-\sqrt{3}-2A)\lambda-\left(2(\sqrt{2}-1)
+\frac{16}{9}\left(\sqrt{6}-\sqrt{3}-\frac{\sqrt{2}}{8}\zeta\right)A
-\frac{8}{9}A^2\right)\lambda^2\right)\sin^2\Delta_{21}\nonumber\\
&&-\left(\frac{1}{3}-\frac{4}{3}(\sqrt{6}-\sqrt{3})\lambda+\left(54-38\sqrt{2}
+\frac{2\sqrt{2}}{3}\zeta A-\frac{4}{3}A^2\right)\lambda^2\right)\sin^2\Delta_{31}\nonumber\\
&&-\left(\frac{2}{3}+\frac{4}{3}(\sqrt{6}-\sqrt{3})\lambda-\left(54-38\sqrt{2}
+\frac{2\sqrt{2}}{3}\zeta
A+\frac{8}{3}A^2\right)\lambda^2\right)\sin^2\Delta_{32},\nonumber\\
P(\nu_\mu\rightarrow\nu_\tau)&=&-\left(\frac{2}{9}-\frac{4}{9}(\sqrt{6}-\sqrt{3})\lambda
-\left(2\sqrt{2}-2+\frac{8}{9}A^2\right)\lambda^2\right)\sin^2\Delta_{21}\nonumber\\
&&+\left(\frac{1}{3}-\frac{4}{3}(\sqrt{6}-\sqrt{3})\lambda+\left(54-38\sqrt{2}
-\frac{4}{3}A^2\right)\lambda^2\right)\sin^2\Delta_{31}\nonumber\\
&&+\left(\frac{2}{3}+\frac{4}{3}(\sqrt{6}-\sqrt{3})\lambda-\left(54-38\sqrt{2}
+\frac{8}{3}A^2\right)\lambda^2\right)\sin^2\Delta_{32}\nonumber\\
&&+\frac{\sqrt{2}}{3}\xi
A\lambda^2(\sin2\Delta_{21}-\sin2\Delta_{31}+\sin2\Delta_{32}),\nonumber
\end{eqnarray}
\begin{eqnarray}
P(\nu_\tau\rightarrow\nu_e)&=&\left(\frac{4}{9}-\frac{8}{9}(\sqrt{6}-\sqrt{3}+A)\lambda-\left(4(\sqrt{2}-1)
-\frac{16}{9}\left(\sqrt{6}-\sqrt{3}-\frac{\sqrt{2}}{8}\zeta\right)A\right)\lambda^2\right)\sin^2\Delta_{21}\nonumber\\
&&-\frac{2\sqrt{2}}{3}\zeta
A\lambda^2(\sin^2\Delta_{31}-\sin^2\Delta_{32})
+\frac{\sqrt{2}}{3}\xi
A\lambda^2(\sin2\Delta_{21}-\sin2\Delta_{31}+\sin2\Delta_{32}),\nonumber\\
P(\nu_\tau\rightarrow\nu_\mu)&=&-\left(\frac{2}{9}-\frac{4}{9}(\sqrt{6}-\sqrt{3})\lambda
-\left(2\sqrt{2}-2+\frac{8}{9}A^2\right)\lambda^2\right)\sin^2\Delta_{21}\nonumber\\
&&+\left(\frac{1}{3}-\frac{4}{3}(\sqrt{6}-\sqrt{3})\lambda+\left(54-38\sqrt{2}
-\frac{4}{3}A^2\right)\lambda^2\right)\sin^2\Delta_{31}\nonumber\\
&&+\left(\frac{2}{3}+\frac{4}{3}(\sqrt{6}-\sqrt{3})\lambda-\left(54-38\sqrt{2}
+\frac{8}{3}A^2\right)\lambda^2\right)\sin^2\Delta_{32}\nonumber\\
&&-\frac{\sqrt{2}}{3}\xi A\lambda^2(\sin2\Delta_{21}-\sin2\Delta_{31}+\sin2\Delta_{32}),\nonumber\\
P(\nu_\tau\rightarrow\nu_\tau)&=&1-\left(\frac{2}{9}-\frac{4}{9}(\sqrt{6}-\sqrt{3}+2A)\lambda-\left(2(\sqrt{2}-1)
-\frac{16}{9}\left(\sqrt{6}-\sqrt{3}-\frac{\sqrt{2}}{8}\zeta\right)A
-\frac{8}{9}A^2\right)\lambda^2\right)\sin^2\Delta_{21}\nonumber\\
&&-\left(\frac{1}{3}-\frac{4}{3}(\sqrt{6}-\sqrt{3})\lambda+\left(54-38\sqrt{2}
-\frac{2\sqrt{2}}{3}\zeta A-\frac{4}{3}A^2\right)\lambda^2\right)\sin^2\Delta_{31}\nonumber\\
&&-\left(\frac{2}{3}+\frac{4}{3}(\sqrt{6}-\sqrt{3})\lambda-\left(54-38\sqrt{2}
-\frac{2\sqrt{2}}{3}\zeta
A+\frac{8}{3}A^2\right)\lambda^2\right)\sin^2\Delta_{32}.\nonumber
\end{eqnarray}

In {\it case 2},
\begin{eqnarray}
P(\nu_e\rightarrow\nu_\mu)&=&\Bigg(\frac{4}{9}-\frac{8}{9}\left(\sqrt{6}-\sqrt{3}
-\left(1+\frac{\sqrt{2}}{4}\zeta'\right)A\right)\lambda\nonumber\\
&&-\left(4(\sqrt{2}-1)+\frac{16}{9}(\sqrt{6}-\sqrt{3})\left(1-\frac{7\sqrt{2}}{8}\zeta'\right)A
+\frac{8}{9}(\zeta'^2+\xi'^2)A^2\right)\lambda^2\Bigg)\sin^2\Delta_{21}\nonumber\\
&&+\left(\frac{2\sqrt{2}}{3}\zeta'A\lambda-\left(\frac{2\sqrt{2}}{3}(\sqrt{6}-\sqrt{3})\zeta'A
-\frac{4}{3}(\zeta'^2+\xi'^2)A^2\right)\lambda^2\right)\sin^2\Delta_{31}\nonumber\\
&&-\left(\frac{2\sqrt{2}}{3}\zeta'A\lambda-\left(\frac{2\sqrt{2}}{3}(\sqrt{6}-\sqrt{3})\zeta'A
+\frac{2}{3}(\zeta'^2+\xi'^2)A^2\right)\lambda^2\right)\sin^2\Delta_{32}\nonumber\\
&&+\frac{\sqrt{2}}{3}\left(\xi'A\lambda-(\sqrt{6}-\sqrt{3})\xi'A\lambda^2\right)
(\sin2\Delta_{21}-\sin2\Delta_{31}+\sin2\Delta_{32}),\nonumber\\
P(\nu_e\rightarrow\nu_\tau)&=&\Bigg(\frac{4}{9}-\frac{8}{9}\left(\sqrt{6}-\sqrt{3}
+\left(1+\frac{\sqrt{2}}{4}\zeta'\right)A\right)\lambda\nonumber\\
&&-\left(4(\sqrt{2}-1)-\frac{16}{9}(\sqrt{6}-\sqrt{3})\left(1-\frac{7\sqrt{2}}{8}\zeta'\right)A
+\frac{8}{9}(\zeta'^2+\xi'^2)A^2\right)\lambda^2\Bigg)\sin^2\Delta_{21}\nonumber\\
&&-\left(\frac{2\sqrt{2}}{3}\zeta'A\lambda-\left(\frac{2\sqrt{2}}{3}(\sqrt{6}-\sqrt{3})\zeta'A
+\frac{4}{3}(\zeta'^2+\xi'^2)A^2\right)\lambda^2\right)\sin^2\Delta_{31}\nonumber\\
&&+\left(\frac{2\sqrt{2}}{3}\zeta'A\lambda-\left(\frac{2\sqrt{2}}{3}(\sqrt{6}-\sqrt{3})\zeta'A
-\frac{2}{3}(\zeta'^2+\xi'^2)A^2\right)\lambda^2\right)\sin^2\Delta_{32}\nonumber\\
&&-\frac{\sqrt{2}}{3}\left(\xi'A\lambda-(\sqrt{6}-\sqrt{3})\xi'A\lambda^2\right)
(\sin2\Delta_{21}-\sin2\Delta_{31}+\sin2\Delta_{32}),\nonumber
\end{eqnarray}
\begin{eqnarray}
P(\nu_\mu\rightarrow\nu_e)&=&\Bigg(\frac{4}{9}-\frac{8}{9}\left(\sqrt{6}-\sqrt{3}-
\left(1+\frac{\sqrt{2}}{4}\zeta'\right)A\right)\lambda\nonumber\\
&&-\left(4(\sqrt{2}-1)+\frac{16}{9}(\sqrt{6}-\sqrt{3})\left(1-\frac{7\sqrt{2}}{8}\zeta'\right)A
+\frac{8}{9}(\zeta'^2+\xi'^2)A^2\right)\lambda^2\Bigg)\sin^2\Delta_{21}\nonumber\\
&&+\left(\frac{2\sqrt{2}}{3}\zeta'A\lambda-\left(\frac{2\sqrt{2}}{3}(\sqrt{6}-\sqrt{3})\zeta'A
-\frac{4}{3}(\zeta'^2+\xi'^2)A^2\right)\lambda^2\right)\sin^2\Delta_{31}\nonumber\\
&&-\left(\frac{2\sqrt{2}}{3}\zeta'A\lambda-\left(\frac{2\sqrt{2}}{3}(\sqrt{6}-\sqrt{3})\zeta'A
+\frac{2}{3}(\zeta'^2+\xi'^2)A^2\right)\lambda^2\right)\sin^2\Delta_{32}\nonumber\\
&&-\frac{\sqrt{2}}{3}\left(\xi'A\lambda-(\sqrt{6}-\sqrt{3})\xi'A\lambda^2\right)
(\sin2\Delta_{21}-\sin2\Delta_{31}+\sin2\Delta_{32}),\nonumber\\
P(\nu_\mu\rightarrow\nu_\mu)&=&1-\Bigg(\frac{2}{9}-\frac{4}{9}\left(\sqrt{6}-\sqrt{3}-
\left(2+\frac{\sqrt{2}}{2}\zeta'\right)A\right)\lambda\nonumber\\
&&-\left(2(\sqrt{2}-1)+\frac{16}{9}(\sqrt{6}-\sqrt{3})\left(1-\frac{7\sqrt{2}}{8}\zeta'\right)A
-\frac{8}{9}\left(1+\frac{\sqrt{2}}{2}\zeta'-\frac{3}{8}\zeta'^2
+\frac{5}{8}\xi'^2\right)A^2\right)\lambda^2\Bigg)\sin^2\Delta_{21}\nonumber\\
&&-\Bigg(\frac{1}{3}-\frac{4}{3}\left(\sqrt{6}-\sqrt{3}-\frac{\sqrt{2}}{2}\zeta'A\right)\lambda\nonumber\\
&&+\left(54-38\sqrt{2}-\frac{2\sqrt{2}}{3}(\sqrt{6}-\sqrt{3})\zeta'A
-\frac{4}{3}\left(1+\sqrt{2}\zeta'-\frac{1}{4}\zeta'^2
-\frac{1}{4}\xi'^2\right)A^2\right)\lambda^2\Bigg)\sin^2\Delta_{31}\nonumber\\
&&-\Bigg(\frac{2}{3}+\frac{4}{3}\left(\sqrt{6}-\sqrt{3}-\frac{\sqrt{2}}{2}\zeta'A\right)\lambda\nonumber\\
&&-\left(54-38\sqrt{2}-\frac{2\sqrt{2}}{3}(\sqrt{6}-\sqrt{3})\zeta'A
+\frac{8}{3}\left(1-\frac{\sqrt{2}}{2}\zeta'+\frac{1}{8}\zeta'^2
+\frac{1}{8}\xi'^2\right)A^2\right)\lambda^2\Bigg)\sin^2\Delta_{32},\nonumber\\
P(\nu_\mu\rightarrow\nu_\tau)&=&-\left(\frac{2}{9}-\frac{4}{9}(\sqrt{6}-\sqrt{3})\lambda
-\left(2(\sqrt{2}-1)+\frac{8}{9}\left(1+\frac{\sqrt{2}}{2}\zeta'
+\frac{5}{8}\zeta'^2+\frac{13}{8}\xi'^2\right)A^2\right)\lambda^2\right)\sin^2\Delta_{21}\nonumber\\
&&+\left(\frac{1}{3}-\frac{4}{3}(\sqrt{6}-\sqrt{3})\lambda+\left(54-38\sqrt{2}
-\frac{4}{3}\left(1+\sqrt{2}\zeta'+\frac{3}{4}\zeta'^2
+\frac{3}{4}\xi'^2\right)A^2\right)\lambda^2\right)\sin^2\Delta_{31}\nonumber\\
&&+\left(\frac{2}{3}+\frac{4}{3}(\sqrt{6}-\sqrt{3})\lambda-\left(54-38\sqrt{2}
+\frac{8}{3}\left(1-\frac{\sqrt{2}}{2}\zeta'+\frac{3}{8}\zeta'^2
+\frac{3}{8}\xi'^2\right)A^2\right)\lambda^2\right)\sin^2\Delta_{32}\nonumber\\
&&+\frac{\sqrt{2}}{3}\left(\xi'A\lambda-(\sqrt{6}-\sqrt{3})\xi'A\lambda^2\right)
(\sin2\Delta_{21}-\sin2\Delta_{31}+\sin2\Delta_{32}),\nonumber\\
P(\nu_\tau\rightarrow\nu_e)&=&\Bigg(\frac{4}{9}-\frac{8}{9}\left(\sqrt{6}-\sqrt{3}
+\left(1+\frac{\sqrt{2}}{4}\zeta'\right)A\right)\lambda\nonumber\\
&&-\left(4(\sqrt{2}-1)-\frac{16}{9}(\sqrt{6}-\sqrt{3})\left(1-\frac{7\sqrt{2}}{8}\zeta'\right)A
+\frac{8}{9}(\zeta'^2+\xi'^2)A^2\right)\lambda^2\Bigg)\sin^2\Delta_{21}\nonumber\\
&&-\left(\frac{2\sqrt{2}}{3}\zeta'A\lambda-\left(\frac{2\sqrt{2}}{3}(\sqrt{6}-\sqrt{3})\zeta'A
+\frac{4}{3}(\zeta'^2+\xi'^2)A^2\right)\lambda^2\right)\sin^2\Delta_{31}\nonumber\\
&&+\left(\frac{2\sqrt{2}}{3}\zeta'A\lambda-\left(\frac{2\sqrt{2}}{3}(\sqrt{6}-\sqrt{3})\zeta'A
-\frac{2}{3}(\zeta'^2+\xi'^2)A^2\right)\lambda^2\right)\sin^2\Delta_{32}\nonumber\\
&&+\frac{\sqrt{2}}{3}\left(\xi'A\lambda-(\sqrt{6}-\sqrt{3})\xi'A\lambda^2\right)
(\sin2\Delta_{21}-\sin2\Delta_{31}+\sin2\Delta_{32}),\nonumber
\end{eqnarray}
\begin{eqnarray}
P(\nu_\tau\rightarrow\nu_\mu)&=&-\left(\frac{2}{9}-\frac{4}{9}(\sqrt{6}-\sqrt{3})\lambda
-\left(2(\sqrt{2}-1)+\frac{8}{9}\left(1+\frac{\sqrt{2}}{2}\zeta'
+\frac{5}{8}\zeta'^2+\frac{13}{8}\xi'^2\right)A^2\right)\lambda^2\right)\sin^2\Delta_{21}\nonumber\\
&&+\left(\frac{1}{3}-\frac{4}{3}(\sqrt{6}-\sqrt{3})\lambda+\left(54-38\sqrt{2}
-\frac{4}{3}\left(1+\sqrt{2}\zeta'+\frac{3}{4}\zeta'^2
+\frac{3}{4}\xi'^2\right)A^2\right)\lambda^2\right)\sin^2\Delta_{31}\nonumber\\
&&+\left(\frac{2}{3}+\frac{4}{3}(\sqrt{6}-\sqrt{3})\lambda-\left(54-38\sqrt{2}
+\frac{8}{3}\left(1-\frac{\sqrt{2}}{2}\zeta'+\frac{3}{8}\zeta'^2
+\frac{3}{8}\xi'^2\right)A^2\right)\lambda^2\right)\sin^2\Delta_{32}\nonumber\\
&&-\frac{\sqrt{2}}{3}\left(\xi'A\lambda-(\sqrt{6}-\sqrt{3})\xi'A\lambda^2\right)
(\sin2\Delta_{21}-\sin2\Delta_{31}+\sin2\Delta_{32}),\nonumber\\
P(\nu_\tau\rightarrow\nu_\tau)&=&1-\Bigg(\frac{2}{9}-\frac{4}{9}\left(\sqrt{6}-\sqrt{3}
+\left(2+\frac{\sqrt{2}}{2}\zeta'\right)A\right)\lambda\nonumber\\
&&-\left(2(\sqrt{2}-1)-\frac{16}{9}(\sqrt{6}-\sqrt{3})\left(1-\frac{7\sqrt{2}}{8}\zeta'\right)A
-\frac{8}{9}\left(1+\frac{\sqrt{2}}{2}\zeta'-\frac{3}{8}\zeta'^2
+\frac{5}{8}\xi'^2\right)A^2\right)\lambda^2\Bigg)\sin^2\Delta_{21}\nonumber\\
&&-\Bigg(\frac{1}{3}-\frac{4}{3}\left(\sqrt{6}-\sqrt{3}+\frac{\sqrt{2}}{2}\zeta'A\right)\lambda\nonumber\\
&&+\left(54-38\sqrt{2}+\frac{2\sqrt{2}}{3}(\sqrt{6}-\sqrt{3})\zeta'A
-\frac{4}{3}\left(1+\sqrt{2}\zeta'-\frac{1}{4}\zeta'^2
-\frac{1}{4}\xi'^2\right)A^2\right)\lambda^2\Bigg)\sin^2\Delta_{31}\nonumber\\
&&-\Bigg(\frac{2}{3}+\frac{4}{3}\left(\sqrt{6}-\sqrt{3}+\frac{\sqrt{2}}{2}\zeta'A\right)\lambda\nonumber\\
&&-\left(54-38\sqrt{2}+\frac{2\sqrt{2}}{3}(\sqrt{6}-\sqrt{3})\zeta'A
+\frac{8}{3}\left(1-\frac{\sqrt{2}}{2}\zeta'+\frac{1}{8}\zeta'^2
+\frac{1}{8}\xi'^2\right)A^2\right)\lambda^2\Bigg)\sin^2\Delta_{32}.\nonumber
\end{eqnarray}
\end{widetext}


\begin{thebibliography}{99}

\bibitem{cabibbo}
N.~Cabibbo, Phys. Rev. Lett. {\bf 10}, 531 (1963).

\bibitem{km}
M.~Kobayashi and T.~Maskawa, Prog. Theor. Phys. {\bf 49}, 652
(1973).

\bibitem{pontecorvo}
B.~Pontecorvo, Sov. Phys. JETP {\bf 6}, 429 (1957).

\bibitem{mns}
Z.~Maki, M.~Nakagawa, and S.~Sakata, Prog. Theor. Phys. {\bf 28},
870 (1962).

\bibitem{ck}
L.L.~Chau and W.Y.~Keung, Phys. Rev. Lett. {\bf 53}, 1802 (1984).

\bibitem{yao}
W.-M.~Yao {\it et al.}, J. Phys. {\bf G33}, 1 (2006).

\bibitem{wolfensteinpara}
L.~Wolfenstein, Phys. Rev. Lett. {\bf 51}, 1945 (1983).

\bibitem{kamland}
KamLAND Collaboration, T.~Araki {\it et al.}, Phys. Rev. Lett. {\bf
94}, 081801 (2005).

\bibitem{sno}
SNO Collaboration, B.~Aharmim {\it et al.}, Phys. Rev. {\bf C72}
055502 (2005).

\bibitem{kamiokande}
Super-Kamiokande Collaboration, Y.~Ashie {\it et al.}, Phys. Rev.
{\bf D71}, 112005 (2005).

\bibitem{minos}
MINOS Collaboration, D.G.~Michael {\it et al.}, Phys. Rev. Lett.
{\bf 97}, 191801 (2006).

\bibitem{k2k}
K2K Collaboration, M.H.~Ahn {\it et al.}, Phys. Rev. {\bf D74},
072003 (2006).

\bibitem{chooz}
CHOOZ Collaboration, M.~Apollonio {\it et al.}, Phys. Lett. {\bf
B466}, 415 (1999); Eur. Phys. J. {\bf C27} 331 (2003); Palo Verde
Collaboration, F.~Boehm {\it et al.}, Phys. Rev. {\bf D64} 112001
(2001).

\bibitem{maltoni}
M.~Maltoni, T.~Schwetz, M.A.~Tortola and J.W.F.~Valle,
hep-ph/0405172v6, New J. Phys. {\bf 6}, 122 (2004).

\bibitem{murayama}
\texttt{http://hitoshi.berkeley.edu/neutrino/}.

\bibitem{gonzalez}
M.C.~Gonzalez-Garcia, M.~Maltoni, arXiv: 0704.1800 [hep-ph].

\bibitem{rodejohann}
W.~Rodejohann, Phys. Rev. {\bf D69}, 033005 (2004); N.~Li and
B.-Q.~Ma, Phys. Lett. {\bf B600}, 248 (2004).

\bibitem{wolfenstein2}
L.~Wolfenstein, Phys. Rev. {\bf D18}, 958 (1978).

\bibitem{harrison}
P.F.~Harrison, D.H.~Perkins and W.G.~Scott, Phys. Lett. {\bf B458},
79 (1999); Phys. Lett. {\bf B530}, 167 (2002); Z.Z.~Xing, Phys.
Lett. {\bf B533}, 85 (2002); P.F.~Harrison and W.G.~Scott, Phys.
Lett. {\bf B535}, 163 (2002); Phys. Lett. {\bf B557}, 76 (2003);
X.G.~He and A.~Zee, Phys. Lett. {\bf B560}, 87 (2003); C.I.~Low and
R.R.~Volkas, Phys. Rev. {\bf D68}, 033007 (2003); A.~Zee, Phys. Rev.
{\bf D68}, 093002 (2003); G.~Altarelli and F.~Feruglio, Nucl. Phys.
{\bf B720}, 64 (2005); Nucl. Phys. {\bf B741}, 215 (2006);
F.~Plentinger and W.~Rodejohann, Phys. Lett. {\bf B625}, 264 (2005);
R.~Friedberg and T.D.~Lee, hep-ph/0606071; R.N.~Mohapatra, S.~Nasri
and H.B.~Yu, Phys. Lett. {\bf B639}, 318 (2006); X.G.~He, Y.Y.~Keum
and R.R.~Volkas, JHEP {\bf 0604}, 039 (2006); G.~Altarelli,
F.~Feruglio and Y.~Lin, Nucl. Phys. {\bf B775}, 31 (2007);
F.~Feruglio, C.~Hagedorn, Y.~Lin and L.~Merlo, Nucl. Phys. {\bf
B775}, 120 (2007); Y.~Koide, J. Phys. {\bf G34}, 1653 (2007); I.~de
Medeiros Varzielas, S.F.~King and G.G.~Ross, Phys. Lett. {\bf B644},
153 (2007); Phys. Lett. {\bf B648}, 201 (2007); C.~Luhn, S.~Nasri
and P.~Ramond, Phys. Lett. {\bf B652} 27 (2007); C.S.~Lam, Phys.
Lett. {\bf B656}, 193 (2007).

\bibitem{li}
N.~Li and B.-Q.~Ma, Phys. Rev. {\bf D71}, 017302 (2005);
J.D.~Bjorken, P.F.~Harrison and W.G.~Scott, Phys. Rev. {\bf D74},
073012 (2006); S.F.~King, Phys. Lett. {\bf B659}, 244 (2008);
S.~Pakvasa, W.~Rodejohann and T.J.~Weiler, Phys. Rev. Lett. {\bf
100}, 111801 (2008).

\bibitem{smirnov}
A.Y.~Smirnov, hep-ph/0402264.

\bibitem{qlc}
H.~Minakata and A.Y.~Smirnov, Phys. Rev. {\bf D70}, 073009 (2004);
M.~Raidal, Phys. Rev. Lett. {\bf 93}, 161801 (2004); P.H.~Frampton
and R.N.~Mohapatra, JHEP {\bf 0501}, 025 (2005); J.~Ferrandis and
S.~Pakvasa, Phys. Rev. {\bf D71}, 033004 (2005); S.K.~Kang, C.S.~Kim
and J.~Lee, Phys. Lett. {\bf B619}, 129 (2005); S.~Antusch,
S.F.~King and R.N.~Mohapatra, Phys. Lett. {\bf B618}, 150 (2005);
M.A.~Schmidt and A.Y.~Smirnov, Phys. Rev. {\bf D74}, 113003 (2006);
F.~Plentinger, G.~Seidl and W.~Winter, Phys. Rev. {\bf D76}, 113003
(2007).

\bibitem{li2}
N.~Li and B.-Q.~Ma, Phys. Rev. {\bf D71}, 097301 (2005).

\bibitem{jarlskog} C.~Jarlskog, Phys. Rev. Lett. {\bf 55}, 1039
(1985); Z. Phys. {\bf C29}, 491 (1985)

\bibitem{farzan}
Y.~Farzan and A.Y.~Smirnov, Phys. Rev. {\bf D65}, 113001 (2002);
Z.Z.~Xing and H.~Zhang, Phys. Lett. {\bf B618}, 131 (2005); H.~Zhang
and Z.Z.~Xing, Eur. Phys. J. {\bf C41}, 143 (2005).


\end{thebibliography}
\end{document}